\documentclass[runningheads]{llncs}
\usepackage{graphicx}
\usepackage{cite}
\usepackage{amsmath,amssymb,amsfonts}
\usepackage{algorithmic}
\usepackage{textcomp}
\usepackage{xcolor}
\usepackage{caption}
\usepackage{subcaption}
\usepackage{float}

\usepackage{listings}
\usepackage{courier}

\definecolor{fixmeColor}{rgb}{1.0,0.2,0.2}
\definecolor{mygreen}{rgb}{0,0.6,0}
\definecolor{mygray}{rgb}{0.5,0.5,0.5}
\definecolor{mymauve}{rgb}{0.58,0,0.82}
\definecolor{darkgray}{rgb}{.4,.4,.4}
\definecolor{purple}{rgb}{0.65, 0.12, 0.82}

\lstdefinelanguage{LF}{
	alsoletter=-,
	keywords={at, deadline, after, state, logical, physical, startup, shutdown,
		reaction, preamble, target, reactor, input, output, constructor, new, timeout, coordination,
		action, comm-type,
		actor, handler, main, federated, serializer, STP, timer, sec,
		secs, msec, msecs, usec, usecs},
	emph={L, trigger, 
		class, name, init, effect, instance,  delay, 
	}, emphstyle=\itshape,
	keywordstyle=\color{black}\bfseries,
	ndkeywords={class, export, boolean, throw, implements, import, this, if, else, time, int, string},
	ndkeywordstyle=\color{darkgray}\bfseries,
	identifierstyle=\color{black},
	sensitive=false,
	comment=[l]{//},
	morecomment=[s]{/*}{*/},
	commentstyle=\color{purple}\ttfamily,
	stringstyle=\color{black}\ttfamily,
	morestring=[b]',
	morestring=[b]"
	}
\begin{document}
%

\title{%
	\vspace{-4em}
			\begin{center}
				\vspace{-4em}
						\footnotesize This is an authors' copy of the paper to appear in Proceedings of the 2nd EAI International Conference on Security and Privacy in Cyber-Physical Systems and Smart Vehicles (SmartSP 2024)
					\end{center}
			A Case Study of API Design for Interoperability and Security of the Internet of Things
			\vspace{-0.5em}
	}

%
\titlerunning{A Case Study of API Design for Interoperability and Security of the IoT}
\author{
	Dongha Kim
\orcidID{0000-0001-7660-9646} 
	\and
Chanhee Lee
\orcidID{0009-0009-0191-6874}
\and
Hokeun Kim
\orcidID{0000-0003-1450-5248}
}
\authorrunning{D. Kim et al.}
\institute{Arizona State University, Tempe, AZ 85281, USA
\email{\{dongha,chanheel,hokeun\}@asu.edu}
\vspace{-2em}
}

\maketitle              
\begin{abstract}
Heterogeneous distributed systems, including the Internet of Things (IoT) or distributed cyber-physical systems (CPS), often suffer a lack of interoperability and security, which hinders the wider deployment of such systems.
Specifically, the different levels of security requirements and the heterogeneity in terms of communication models, for instance, point-to-point vs. publish-subscribe, are the example challenges of IoT and distributed CPS consisting of heterogeneous devices and applications.
In this paper, we propose a working application programming interface (API) and runtime to enhance interoperability and security while addressing the challenges that stem from the heterogeneity in the IoT and distributed CPS.
In our case study, we design and implement our application programming interface (API) design approach using open-source software, and with our working implementation, we evaluate the effectiveness of our proposed approach.
Our experimental results suggest that our approach can achieve both interoperability and security in the IoT and distributed CPS with a reasonably small overhead and better-managed software.

\keywords{
	Internet of Things
	\and
	Interoperability
	\and
	Security
	\and
	API Design.
	
	}
\end{abstract}
\section{Introduction}

Heterogeneous distributed systems, including the Internet of Things (IoT) or distributed cyber-physical systems (CPS), have been rising, especially with the benefits of edge computing, such as low latency, privacy protection, and scalability~\cite{yu2017survey,ning2020heterogeneous}.
As more IoT and distributed systems running on the edge are used in many different domains, it is increasingly challenging to support a diversity of communication models because of the heterogeneity in edge-computing environments where the heterogeneous devices and applications interact with one another~\cite{carvalho2021edge}.
One representative such application is the smart city applications~\cite{costin2019need, koo2021interoperability} that work across the boundaries of different domains, including smart homes and buildings with various purposes under heterogeneous environments.

Thus, the problem of \emph{interoperability}~\cite{sami2022interoperability} between various components in the IoT and distributed CPS has become one of the major obstacles blocking further deployment of such systems.
This interoperability problem includes dealing with
the different security requirements depending on the specific applications and the characteristics of their data~\cite{xiao2019edge, zeyu2020survey}.
For example, safety-critical applications such as transportation or medical applications will require stronger security guarantees~\cite{tedeschi2019edge}, while environmental sensing systems may require minimal security, such as data integrity while prioritizing low power consumption over security~\cite{shapsough2018securing}.

There have been research efforts to address the interoperability and security problems in the IoT and distributed CPS, such as a semantic-based collaboration platform~\cite{sigwele2018intelligent}, networking strategies for interoperability in real-time systems~\cite{gomez2023strategies}, or a secure network for mobile edge computing~\cite{lai2021secure}.
As programming models and APIs play a critical role in the development and deployment of edge-based IoT and CPS~\cite{giang2018fog, li2021edge}, programming models dedicated to edge-based systems~\cite{lin2020survey} have been proposed.
However, to the best of our knowledge, there has not been enough research work on programming models or APIs tailored to the IoT and distributed CPS with a working implementation trying to address both interoperability and security problems as a single integrated platform.

In this paper, we design an API and conduct a case study with the working implementation of the runtime based on the designed API to address the interoperability and security issues of the IoT and distributed CPS.
Our contributions are threefold: 
\begin{enumerate}
	\item  We design an API that supports multiple communication models, including point-to-point (e.g., client-server) and publish-subscribe paradigms, to facilitate seamless interaction between heterogeneous devices.
	\item We incorporate a flexible security framework that can be adaptively applied based on varying security requirements.
	\item We develop a working runtime system using open-source platforms to demonstrate the practicality of the API, and evaluate its performance, showing that it achieves interoperability and security with reasonably small overhead while simplifying software development and enhancing maintainability.
\end{enumerate}

\section{Related Work}

Extensive surveys and reviews have been given by the literature 
~\cite{abounassar2022security,lee2021survey,amjad2021systematic,noura2019interoperability,gurdur2018systematic,rana2021systematic} on the interoperability and security issues in the IoT and distributed CPS.
Abounassar \textit{et al.}~\cite{abounassar2022security} point out that the IoT, especially in the healthcare sector, still suffers from security and interoperability challenges even now in the 2020s.
Lee  \textit{et al.}~\cite{lee2021survey} provide a survey on current standards and efforts to bring interoperability and security to the IoT; however, they also present a number of remaining challenges, including the lack of developer support for enabling interoperability and limited security considerations in standards.
A survey by Amjad \textit{et al.}~\cite{amjad2021systematic} reviews the interoperability and security challenges in industrial IoT (IIoT) with a focus on data transfer and application protocols, including the security vulnerabilities in the widely used publish-subscribe protocols~\cite{rana2021systematic}.
Noura \textit{et al.}~\cite{noura2019interoperability} discuss the interoperability issues in the IoT from various perspectives, including, network interoperability, syntactical/semantic interoperability, and platform interoperability.
G{\"u}rd{\"u}r and Asplund~\cite{gurdur2018systematic} emphasize the importance of interoperability in CPS under distributed environments.

There have been several efforts and approaches to dealing with interoperability and security in the IoT and distributed CPS.
Oh \textit{et al.}~\cite{oh2022security} identify the security threats in the widely used interoperability protocol for heterogeneous IoT platforms, the OAuth 2.0 framework.
Margarita \textit{et al.}~\cite{tiziana2021digital} consider inter-operation among various entities, including IoT devices and production lines focused on Industry 4.0; however, this approach primarily targets the REST communication model and does not support various network protocols or security concerns.
Pereira \textit{et al.}~\cite{pedro2023middleware} propose an ontology-based middleware to mitigate interoperability issues, also in industrial IoT (IIoT).
The OPC unified architecture (OPC UA)~\cite{opcua_2008} has been designed to support the interoperability between the smart sensors and the cloud by embracing heterogeneous communication protocols.

Some research work focuses on encryption or authentication to ensure security in the IoT and distributed CPS.
Kannan \textit{et al.}~\cite{b2023military} apply encryption and digital signatures only targeting military applications to secure communication in an IoT environment.
Sensor data, commonly used in military equipment, are collected and tested using encryption to show the effectiveness against data modification.
This literature~\cite{b2023military}, however, does not consider the authentication and authorization process.
Quadir \textit{et al.}~\cite{md2020puf} propose an authentication framework targeting consumer electronics in medical, home, and personal applications.
The framework proposed by Quadir \textit{et al.}~\cite{md2020puf} mainly depends on a centralized server to communicate the distributed device via TLS.
Secure Swarm Programming Platform (SSPP)~\cite{kim2023sspp} is a conceptual platform for providing security at the programming platform level, however, yet without a concrete implementation. 

Also, some prior arts address the security of edge-based IoT and distributed CPS running on Robot Operating System (ROS).
Sanchez \textit{et al.}~\cite{Sanchez2020ros} integrate a smart personal protective system into ROS with edge computing.
For communication between devices, only the MQTT protocol is provided with a Mosquitto message broker, and encryption based on SSL is considered only.
Zhang \textit{et al.}~\cite{Zhang2022ros2} utilize ROS2, where data distribution service (DDS) is used only to provide authentication and access control.
This approach, however, depends on both cloud computing and edges.
It causes user privacy problems immediately across the cloud and edge and can experience performance degradation due to communication overhead. 

\section{Proposed Approach}

This section introduces our proposed API and runtime, designed based on our observation of different modes of communication and security requirements that are common to the IoT and distributed CPS.

\begin{figure}
	\centering
	\includegraphics[width=0.6\columnwidth]{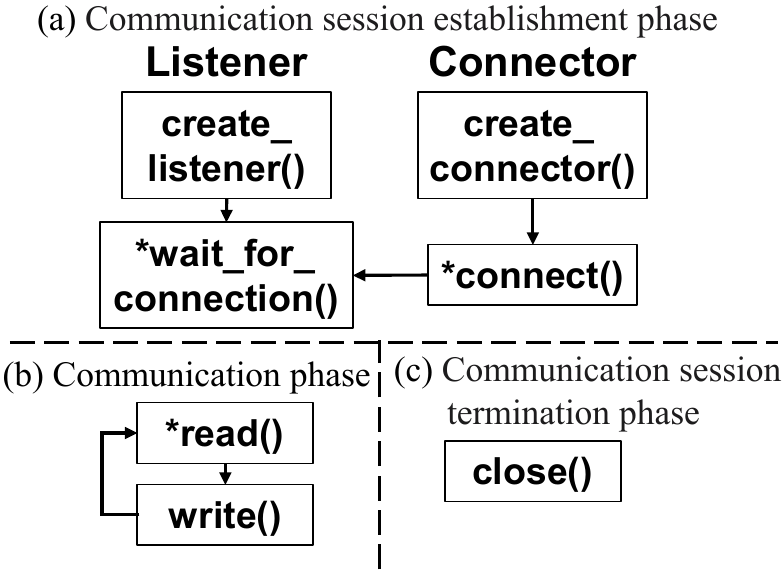}
	\caption{Proposed API functions. Functions with * are blocking functions.}
	\label{fig:APIOverview}
\end{figure}

For the discussion of our API design, we define two types of nodes, i.e., \textit{Listener} and \textit{Connector} (shown in \figurename~\ref{fig:APIOverview}), to represent typical nodes in the IoT and distributed CPS.
We note that these terms (\textit{Listener} and \textit{Connector}) or their concept are not novel, but we use these terms to capture the important and common characteristics of the nodes used in networked and distributed systems, mainly for our discussion in this paper. 
A \textit{Connector} requests a connection to be waited and accepted by a \textit{Listener}.
These nodes are defined so that we can flexibly and seamlessly support different modes of communication and optional security guarantees.

\subsection{Design of API Functions}

Our proposed API consists of seven functions for \textit{Listeners} and \textit{Connectors}. 
These functions are used to establish connections and transfer data between different communication models. 
The API is roughly divided into three phases: communication session establishment phase, communication and data transfer phase, and communication session termination phase.

The first phase involves establishing a single session between two nodes as shown in \figurename~\ref{fig:APIOverview}a. 
A \textit{Connector} and a \textit{Listener} initiate the process by creating their respective objects.
The \textit{Listener} calls \textit{wait\_for\_connection()}, while the \textit{Connector} calls \textit{connect()}, which are both blocking functions, to establish a communication session.
The second phase is for data transfer between the \textit{Listener} and the \textit{Connector} as shown in \figurename~\ref{fig:APIOverview}b. 
For this purpose, we use \textit{read()} and \textit{write()}, which encapsulate the application layer protocol's data transfer.
The final phase is disconnecting the session, using \textit{close()}, as shown in \figurename~\ref{fig:APIOverview}c. 
This \textit{close()} function sends a disconnection request to the other node and releases the resources associated with the session.

\subsection{Interoperability Support}

Our proposed API is designed to be applied to various communication models.
We apply our API to two representative communication models that are most commonly used in the IoT and distributed CPS.

\subsubsection{Point-to-Point Communication}

\begin{figure}
	\centering
	\includegraphics[width=0.88\columnwidth]{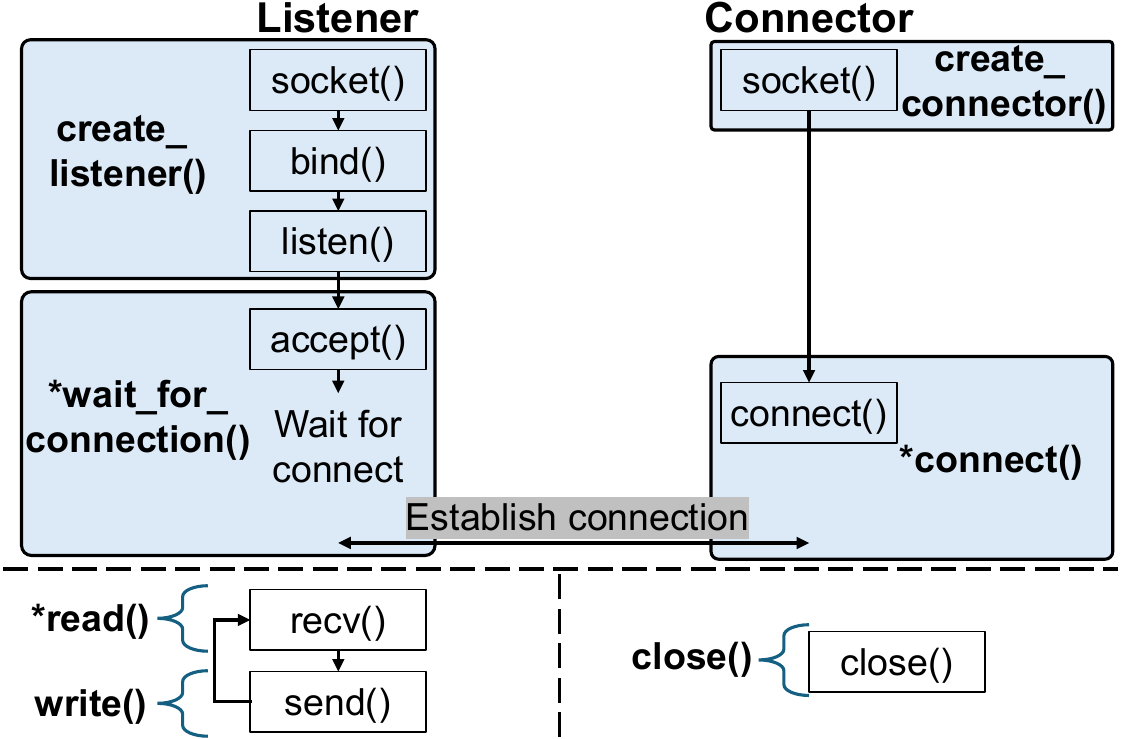}
	\caption{Proposed API functions. Functions with * are blocking functions. These seven functions are used for the following three communication phases of networked nodes: (a) Communication session establishment (initialization) phase, (b) communication (data transfer) phase, and (c) communication session termination (closing) phase.}
	\label{fig:ClientServerOverview}
\end{figure}

\figurename~\ref{fig:ClientServerOverview} shows how our API functions work for point-to-point communication, for example, a client-server model such as TCP or FTP.

Here, we take TCP as an example.
A server and a client in TCP correspond to a \textit{Listener} and a \textit{Connector} in our proposed API, respectively.
On the server side, our API function \textit{create\_listener()} handles the server socket creation, binding, and listening of the socket.
The API function \textit{wait\_for\_connection()} waits and accepts a client's request with blocking. 
On the client side, our API function \textit{create\_connector()} initializes the client socket, and another API function \textit{connect()} calls the TCP socket function, connect(), which establishes a TCP connection with the server.
After the session is set up, API functions \textit{read()} and \textit{write()} perform TCP socket functions recv() and send() to transfer data. 
Finally, \textit{close()} is used to shut down and close the TCP sockets, disconnecting the server and the client.

\subsubsection{Publish-Subscribe Communication}
\label{sec:Publish/Subscribe}

\begin{figure}
	\centering
	\includegraphics[width=0.7\columnwidth]{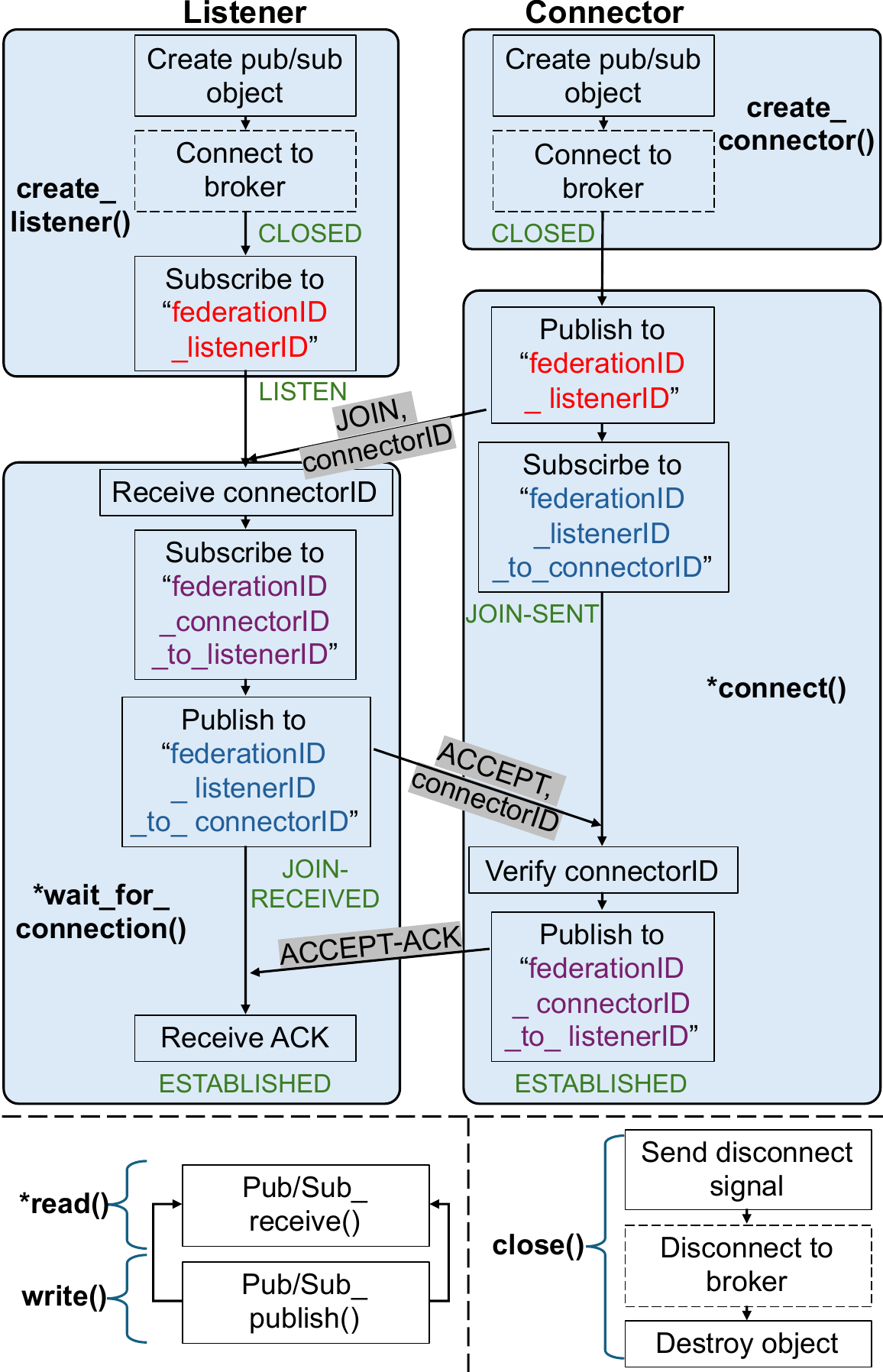}
	\caption{Execution model of the proposed API for publish-subscribe communication. Blocking functions are marked *.}
	\label{fig:PubSubOverview}
\end{figure}

Our proposed API design also supports pub-sub communication as illustrated in \figurename~\ref{fig:PubSubOverview}.
The main challenge in adopting pub-sub into our API is the session establishment.
For the discussion in this paper, we call an application running on nodes of an distributed system running on the edge a \textit{federation}.
We call a unique identifier of a federation \textit{federationID}.
In our API, we assume that all distributed nodes know \textit{federationID} as well as the IP address/port number of the centralized coordinator before the federation starts.
Prior arts such as Robot Operating System 1 (ROS1)~\cite{quigley2009ros} are based on a similar assumption; for example, the ROS Master's IP address and port number are known to all nodes before the runtime starts.

As shown in \figurename~\ref{fig:PubSubOverview}, \textit{create\_listener()} and  \textit{create\_connector()} create \textit{Listener} and \textit{Connector} nodes with an initialized pub/sub object.
If the underlying pub-sub communication uses a centralized third party message broker such as the MQTT~\cite{mqtt} broker, it will also connect to the broker.
It is important to note that this broker is not considered a \textit{Listener} or \textit{Connector} node, as it is a feature of the pub-sub protocol itself rather than part of our design. Therefore, this connection step is bypassed in decentralized pub-sub protocols like Data Distribution Service (DDS)~\cite{omgdds}.
Both \textit{Listener} and \textit{Connector} start from a CLOSED state.
The \textit{Listener} node additionally subscribes to a topic named ``\textit{federationID}\_\textit{listenerID}'' and enters a LISTEN state.

After that, the \textit{Listener} and the \textit{Connector} establish a session through a three-way handshake.
First, the \textit{Connector} node publishes to the topic ``\textit{federationID}\_\textit{listenerID}'', sending a JOIN message with its \textit{connectorID}.
The \textit{listenerID} should be known to the \textit{Connector}, such as the topic name should be known for broadcasting communication.
The \textit{Connector} also subscribes to the topic ``\textit{federationID}\_\textit{listenerID}\_to\_\textit{connectorID}''.
The \textit{Connector} enters a JOIN-SENT state.

The \textit{Listener} node receives the \textit{connectorID}, and subscribes to the topic ``\textit{federationID}\_\textit{connectorID}\_to\_\textit{listenerID}''. 
Then it publishes an ACCEPT message and its \textit{listenerID} to the topic ``\textit{federationID}\_\textit{listenerID}\_to\_\textit{connectorID}''.
The \textit{Listener} enters a JOIN-RECEIVED state.

The \textit{Connector} verifies the received \textit{connectorID}, publishes an ACCEPT-ACK message to ``\textit{federationID}\_\textit{connectorID}\_to\_\textit{listenerID}'' which is the specific topic for the \textit{Listener}, and enters an ESTABLISHED state.
The \textit{Listener} finally receives the ACCEPT-ACK message and also enters an ESTABLISHED state.
Finally, there are two topics for one \textit{Connector}-\textit{Listener} session, which are one-way for each.

The \textit{read()} operation is a synchronous call, causing the node's execution to block until data is received from the corresponding \textit{write()} operation on another node. 
The \textit{close()} sends disconnect signals to the other node, disconnects to the broker if exists, and destroys the pub/sub object.

\subsection{Security Support}

\begin{figure}
	\centering
	\includegraphics[width=0.64\columnwidth]{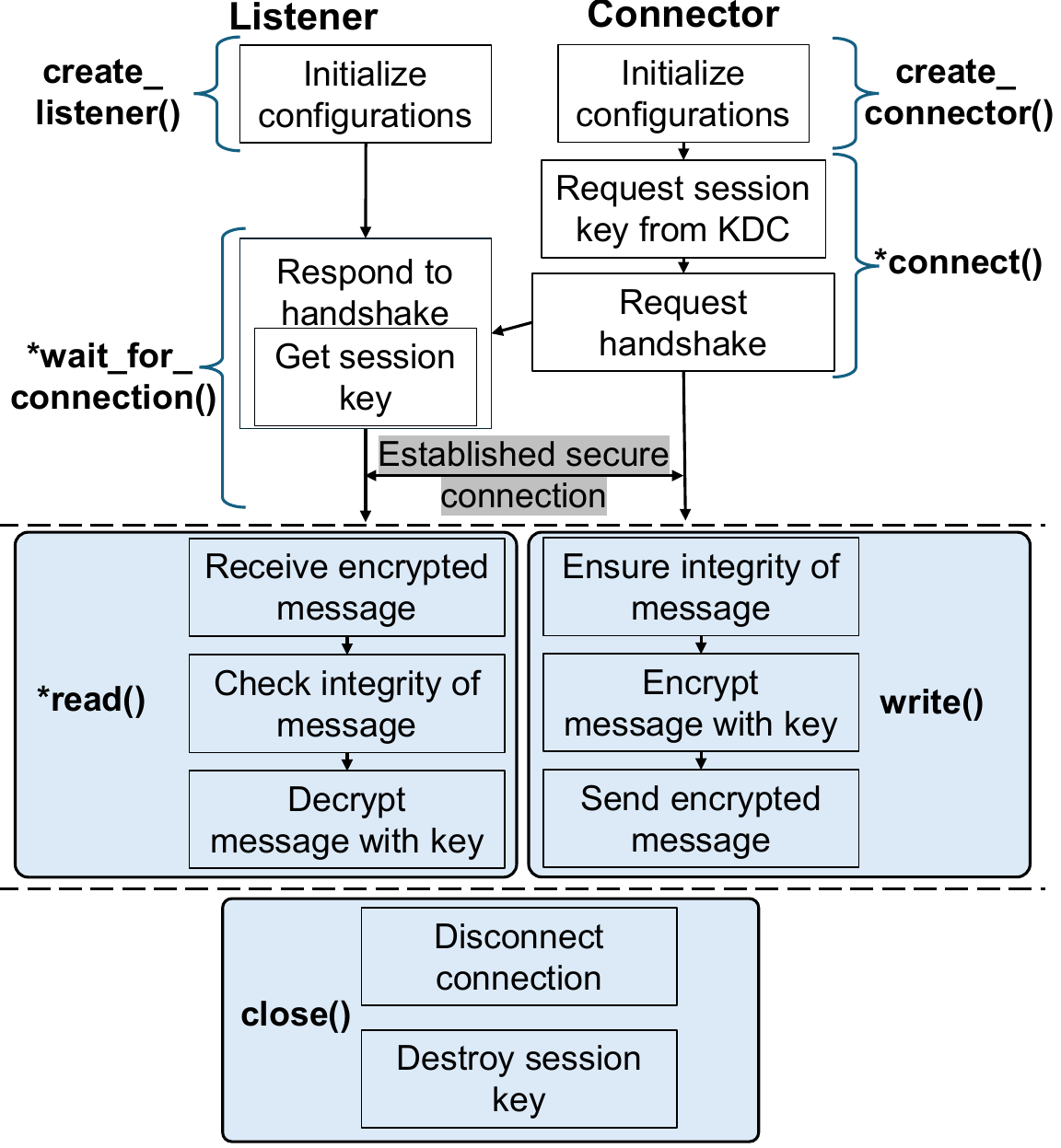}
	\caption{Execution model of the proposed API with security enabled. Functions with * are blocking functions.}
	\label{fig:SecurityOverview}
\end{figure}

For the authentication of each node, we assume that there is a key distribution center (KDC), which is a trusted third party nor Listener or Connector.
KDC is only responsible for generating and distributing encryption keys to multiple \textit{Listeners} and \textit{Connectors} over a network, and do not directly participate in the system's communication.
For example, a well-known network authentication protocol, Kerberos~\cite{saltzer1984end}, uses a centralized authentication server (AS) as its KDC.
The KDC issues tickets containing encrypted session keys, which are symmetric cryptographic keys as a common secret between two nodes, allowing nodes to authenticate and establish secure communication.

\figurename~\ref{fig:SecurityOverview} illustrates an overview of how our API design seamlessly applies security.
The \textit{create\_listener()} and \textit{create\_connector()} first initialize the \textit{Listener} and \textit{Connector} object with configurations, including any certificates necessary for authentication with the KDC.
When \textit{connect()} is invoked, the \textit{Connector} requests a session key from the KDC and sends a connection request to the \textit{Listener}.
Next, the \textit{Listener} will call \textit{wait\_for\_connection()} and wait for a connection request from the \textit{Connector}. 
When the request arrives, the \textit{Listener} must obtain the same session key from the KDC to ensure secure communication with the \textit{Connector}.
The connection request includes a handshake to ensure that both ends have the same session key and session nonce to encrypt the traffic.

When security is enabled, the \textit{write()} function internally encrypts the message to ensure confidentiality and adds authentication methods, such as Message Authentication Codes (MAC), to maintain message integrity.
The \textit{read()} function checks the integrity of received messages through the MAC and decrypts the message content using the session key.
Finally, \textit{close()} disconnects the connection and cleans up the resource related to the secure session, such as the session key.

\section{A Case Study: Design and Implementation}

This section provides a detailed overview of the design and implementation for a case study of the proposed API design, including example code.
For efficient implementation, we leverage open-source software libraries and runtime.

\subsection{Open-Source Software Used for Implementation}

We leverage existing open-source software projects for our case study.
For the communication and coordination among distributed nodes in the IoT or distributed CPS, we use an open-source coordination language and runtime, Lingua Franca (LF)~\cite{lohstroh2021toward}, and for security support, Secure Swarm Toolkit (SST)~\cite{kim_sst_2017}.

\subsubsection{Lingua Franca}
\label{sec:lf_background}

Lingua Franca (LF) is a coordination language designed to guarantee deterministic concurrency with \textit{reactors}~\cite{lohstroh2020reactors}.
Reactors are lightweight concurrent entities that communicate with each other via timestamped messages.
LF's C-runtime supports \textit{federated execution}~\cite{bateni2023risk}, allowing reactors to run across distributed systems and communicate over networks. 
The resulting networked system is called a \textit{federation}, with each individual component known as a \textit{federate}. 
The LF compiler generates separate code for each federate.

LF also provides a separate run-time infrastructure (RTI), which manages time synchronization among federates, ensuring a deterministic message flow and coordinating startup and shutdown. 
RTI also facilitates message exchange among federates, working as a message broker in \textit{centralized coordination}.

\subsubsection{Secure Swarm Toolkit}
The Secure Swarm Toolkit (SST) is an open-source toolkit framework designed to offer robust authorization and authentication mechanisms for distributed environments.
The local entity \textit{Auth}~\cite{kim_secure_2016}, provides authentication and authorization for its locally registered entities, and also supports resilience to migrate trust to another Auth~\cite{kim_resilient_2020}.
SST supports a C API~\cite{kim_sst_2023} designed to support resource-constrained devices.
It has also been used for access control of decentralized and distributed file systems~\cite{jo2023secure}.

\subsection{Software Design Considerations}

We chose LF as the runtime environment of our target systems primarily due to LF's compatibility with a wide range of embedded platforms~\cite{jellum2023beyond}, including Arduino~\cite{arduino}, RP2040~\cite{rp2040}, Zephyr~\cite{zephyr}, and also bare metal devices. 
However, LF has some limitations on network communication, relying on TCP sockets for message exchanges between RTI and federates.
This reliance on TCP restricts interoperability, especially in heterogeneous environments.

For authentication of the federates, LF has two options, basic \textit{federationID} check, and key-hashed message authentication code (key-hashed MAC or HMAC) authentication.
As default, LF currently supports the basic \textit{federationID} check when a federate joins a federation although there is no encryption or message integrity involved. 
A federate sends its \textit{federationID} in plaintext to the RTI, which verifies if the \textit{federationID} is correct.
However, this allows malicious entities to join the federation if they can eavesdrop the \textit{federationID} sent over the network. 
To address this vulnerability, HMAC is used to secure the connection between devices~\cite{kim2023poster}, but LF still does not provide message encryption over the network.

To address the aforementioned security gaps, we integrated SST into LF.
SST is well-suited for large-scale distributed environments, with the Auth providing decentralized authentication and authorization.
SST's flexibility in supporting lightweight cryptography and hash algorithms allows for the customization of security levels to meet different requirements for heterogeneous devices.
The C APIs in SST made integration with LF's C runtime straightforward, thus enhancing both interoperability and security within our distributed system.

\subsection{Code-Generation and Compilation}

\begin{figure}
	\begin{subfigure}[c]{1.0\linewidth}
		\centering
		\includegraphics[width=0.65\textwidth]{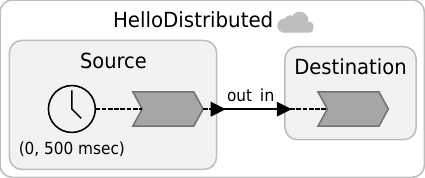}
		\caption{Diagram of a simple federated Lingua Franca program with two federates, \texttt{Source} and \texttt{Destination}.}
		\label{fig:HelloDistributedDiagram}
	\end{subfigure}
	\begin{subfigure}[c]{1.0\linewidth}
		\begin{lstlisting}[language=LF, escapechar=|, label={lst:SimpleFederated}]
			target C {
				coordination: centralized, |\label{centralized}|
				|\textcolor{red}{\textbf{comm-type}: MQTT,}||\label{comm-type}|
				timeout: 500 sec, |\label{timeout}|
				auth: true |\label{auth-true}|
			}
			
			reactor Source {
				output out: int
				timer t(0, 500 msec) |\label{timer}|
				reaction(t) -> out {=
					lf_set(out, 0); |\label{lf_set}|
					=}
			}
			
			reactor Destination {
				input in: int
				reaction(in) {=
					lf_print("Dest received: %s", in->value);
					=}
			}
			
			federated reactor HelloDistributed{
				s = new Source()       
				d = new Destination()  
				s.out -> d.in  |\label{port_connect}|        
		}		\end{lstlisting}
		\vspace{-2\baselineskip}
		\caption{Lingua Franca code for HelloDistributed.lf.}
		\label{fig:HelloDistributedCode}
	\end{subfigure}
	\caption{Example of a simple Lingua Franca program.}
	\label{fig:HelloDistributed}
\end{figure}

To illustrate how we enhance interoperability and security in LF, a simple LF program is organized as shown in \figurename~\ref{fig:HelloDistributed}. 
\figurename~\ref{fig:HelloDistributedDiagram} represents a diagram of the example LF program where an integer value is sent from a \texttt{Source} node to a \texttt{Destination} node. 
Note that the diagram is automatically generated in either Visual Studio Code or Eclipse IDE after LF plug-in is installed.
In \figurename~\ref{fig:HelloDistributedCode}, the LF program named \texttt{HelloDistributed} has two reactors \texttt{Source} and \texttt{Destination}. 
With this LF code, LF compiler generates the executable binaries of two federates for two reactors automatically.

Line~\ref{centralized} indicates that the whole execution of the generated executable binaries is performed with \textit{centralized coordination}, which means that the RTI mediates all messages between the generated federates.
Line~\ref{auth-true} sets the federate joining process to use the HMAC authentication.
With \textit{centralized coordination}, in line~\ref{lf_set}, the message sent from \texttt{Source} reactor is transferred to RTI first and then forwarded to the \texttt{out} port of the \texttt{Destination} reactor.
During communication between federates, the RTI acts like a message broker, providing deterministic timing controls and traceability of all messages among all federates.

In addition to the existing target properties in LF, we newly add a property to select the underlying communication stack as part of our proposed approach.
In line~\ref{comm-type}, the \texttt{comm-type} keyword enables users to specify a network protocol selectively to be used.
Currently, this new feature supports three communication stacks: TCP for client-server, MQTT for pub-sub, and SST for a security model. 
Note that other communication protocols can be easily added to this feature based on our proposed API design.


\begin{figure}
	\centering
	\includegraphics[width=1.0\columnwidth]{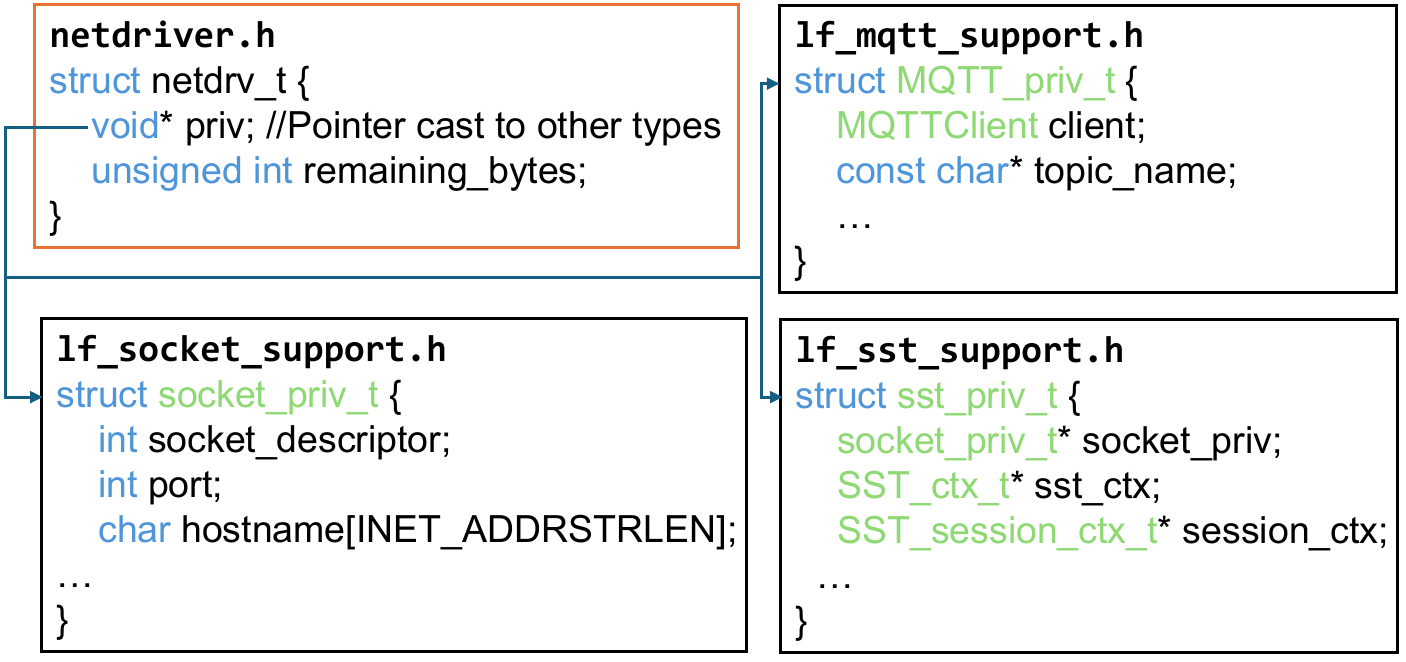}
	\caption{The structure of the \textit{netdriver}. The void pointer in \texttt{netdrv\_t} is cast to another struct depending on the network type for polymorphism.}
	\label{fig:netdriver}
\end{figure}

The LF program is conditionally compiled to ensure federates install only the libraries they need, avoiding unnecessary dependencies.
To enable this flexibility, LF includes a network abstraction layer called the \textit{network driver} (netdriver), which connects low-level communication protocols to the high-level API. 
The netdriver's design allows for polymorphism, enabling it to switch between different communication methods without additional code changes.
As shown in \figurename~\ref{fig:netdriver}, the netdriver contains a void pointer which is cast to another struct pointer depending on the communication type.
This enables providing a unified interface for multiple communication protocols.
The details of \textit{remaining\_bytes} will be explained in section~\ref{sec:read/write}.
To avoid confusion from the POSIX system calls, the defined APIs include a \textit{netdrv\_} prefix, such as \textit{netdrv\_read()}.

\subsection{Runtime Implementation with Proposed API}
We illustrate how our API deisgn supports network interoperability and security.
\subsubsection{Client-Server Runtime}
We implement the client-server model using TCP sockets, which establish connections and enable data transfer between the RTI and federates. 
Since functions of TCP protocol can be easily mapped to the proposed API, the implementation is straightforward.
\subsubsection{Publish-Subscribe Runtime}
We build the pub-sub model based on the MQTT protocol~\cite{mqtt}, utilizing the Eclipse Paho C library~\cite{EclipsePahoC}, a widely used MQTT client library for C.

The proposed three-way handshake in section~\ref{sec:Publish/Subscribe}, establishes a unique session for each connection.
The RTI acts as the \textit{Listener} with a designated \textit{Listener} ID of `RTI', while each federate has a unique \textit{Connector} ID.
Setting the MQTT Quality of Service (QoS) level to 2 ensures that each message is delivered exactly once.
Moreover, synchronous send and receive operations are critical in LF, providing reliability to message exchanges.

\subsubsection{Security}
To improve security, we integrated SST into LF.
When a federate wants to establish a secure session with the RTI, it initiates the connection through \textit{netdrv\_connect()}.
The federate requests for a session key from the \textit{Auth}, which acts as a KDC.
The Auth authenticates the federate through a three way handshake, and then sends out a session key.
The federate now sends a connection request to the RTI sending the received session key's ID.
The RTI receives the request from the federate, and requests the Auth the same session key with the matching key ID.
Once the RTI and federate have the same key, they complete the handshake, establishing a secure session.
Subsequent messages between the federate and the RTI is encrypted and decrypted using the session key.

\subsection{Addressing Further Implementation Challenges}
\label{sec:read/write}

One of the significant challenges in implementation of our proposed API decouples network-related components that are tightly integrated with the LF code base.
Many existing software platforms use a single communication model, thus, their implementation is deeply coupled with the underlying network protocol for various optimizations, for example, using the socket-level buffers for parsing network messages.
Similarly, the current LF implementation is also integrated with TCP.
We discuss how we tackle this problem in our implementation of the proposed API design.

To receive messages, LF uses the POSIX \textit{read()} functions.
However, in some cases, it calls \textit{read()} multiple times to process a single message, first to retrieve the message type and then again to collect the rest of the payload. 
Similarly, LF sometimes sends messages in multiple chunks with separate \textit{write()} calls. 
These inconsistent number of system calls for a single message increases code complexity, requiring additional logic to manage these operations.
Also, it increases the chance to cause I/O errors multiple times and makes it difficult to trace errors.

The proposed \textit{netdrv\_read()} and \textit{netdrv\_write()} are designed to receive and send complete messages.
It is critical to establish a clear mechanism to manage message boundaries and length. 
Protocols like MQTT and SST operate at the application layer and include information on the message length to ensure message boundaries are maintained. 
However, TCP is a stream-based transport layer protocol, meaning data is transmitted as a continuous stream of bytes without explicit message boundaries. 
This characteristic introduces challenges when trying to receive complete messages.

Managing message boundaries and lengths can be straightforward when each message explicitly indicates its total length. 
However, like other existing software platforms, not all LF messages have the payload length in their header.
We classify LF messages into three types: \textit{fixed-length}, \textit{structured variable length}, and \textit{variable length}, which can also be generalized for other distributed-system platforms.

\begin{figure}
	\centering
	\includegraphics[width=1.0\columnwidth]{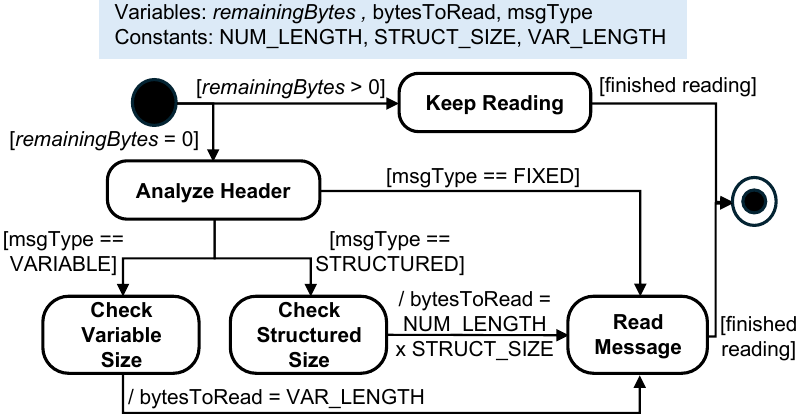}
	\caption{The simplified state machine for the \textit{netdrv\_read()}. The boldface are the states and italic is the input.}
	\label{fig:StateMachine}
\end{figure}

Fixed-length messages have constant-sized payloads as defined by the message type.
For example, MSG\_TYPE\_TIMESTAMP, which is one of the message types in LF for sending the start time of the RTI and federates, has a fixed payload of 8 bytes.
Next, structured variable-length messages have a variable number of fixed-length structures.
So, the message consists of an integer indicating the number of structures of `n,' followed by `n' fixed-length structures.
Finally, the variable length has an integer `n' indicating the length of the payload, followed by the variable payload.

To address this challenge, the TCP model's \textit{netdrv\_read()} uses a state machine to handle each type of message.
\figurename~\ref{fig:StateMachine} shows a simplified version of the state machine.
The \textit{netdrv\_read()} starts with an INITIAL state denoted by a filled black circle. Depending on the size indicated by \textit{remainingBytes}, the state goes to either \texttt{Analyze Header} or \texttt{Keep Reading} state.
When the \textit{netdrv\_read()} function is called and the \textit{remainingBytes} is more than zero, the state machine enters a \texttt{Keep Reading} state.
In this state, \textit{netdrv\_read()} reads as much data as the buffer allows, updating \textit{remainingBytes}.
This design provides flexibility for processing variable-length messages and reduces the risk of buffer overflows.

If the state machine enters an \texttt{Analyze Header} state, it calls the POSIX \textit{read()} to receive only one byte to check a message type.
Regarding the message type, \textit{bytesToRead} is set to indicate the number of bytes of the payload.
If the message type is a fixed-length message, it reads exactly the amount according to the message type header after entering into a \texttt{Read Message} state.
For structured length messages, it calculates \textit{bytesToRead} with \texttt{NUM\_LENGTH} and \texttt{STRUCT\_SIZE}, i.e., the number of structures and a struct size, respectively.
For variable length messages, it first reads the integer `n' indicating the variable length, and then reads `n' bytes.
The netdriver structure keeps track of the additional bytes to be read through an integer, \textit{bytesToRead}, until it reaches to the end state from \texttt{Read Message} state.


\section{Evaluation}

In this section, we evaluate our APl's overhead in terms of communication time, binary size, and message lengths.
We also clarify the strong points of the proposed API design qualitatively through a software design analysis.

\subsection{Experimental Setup}

An IoT or distributed CPS that runs the LF program in \figurename~\ref{fig:HelloDistributed}, where the federate \texttt{Source} sends a message to the federate \texttt{Destination} is used for the evaluation of our API.
The RTI, Mosquitto message broker~\cite{light2017mosquitto} for MQTT, and Auth for SST run on a workstation as the edge, equipped with an i9-13900 CPU and 128 GB memory.
The federates are deployed on two Raspberry Pi 4 (Model B 4GB RAM) devices.
The workstation and the devices communicate over Wi-Fi, and the average end-to-end round-trip latency from the Raspberry Pi to the workstation, measured by a simple ping, is 13.60 milliseconds.
The federates joining the federation use HMAC authentication mode as a baseline.


\subsection{Communication Time Overhead}
\label{sec:comm_overhead}
To estimate the communication time overhead, we measure a \emph{lag}, the time difference between the physical time and logical time. 
Specifically, we measure the time lapse between the system clock (physical time) and the semantic notion of intended global time agreed by all nodes (logical time)~\cite{lohstroh2020language}.
From the example above in \figurename~\ref{fig:HelloDistributedCode}, the timer on line~\ref{timer} triggers the \texttt{Source} reactor to send a message to the \texttt{Destination} reactor every 500 milliseconds.
Due to the timeout in line~\ref{timeout}, this will stop when the logical time reaches 500 seconds.
We also set the timer with a period of 50 milliseconds, with a timeout of 50 seconds, to test sending messages in shorter periods.
Consequently, in both cases, the message will be sent 1,000 times.
We measure the average of each message's lag.

\begin{figure}
	\centering
	\includegraphics[width=0.75\columnwidth]{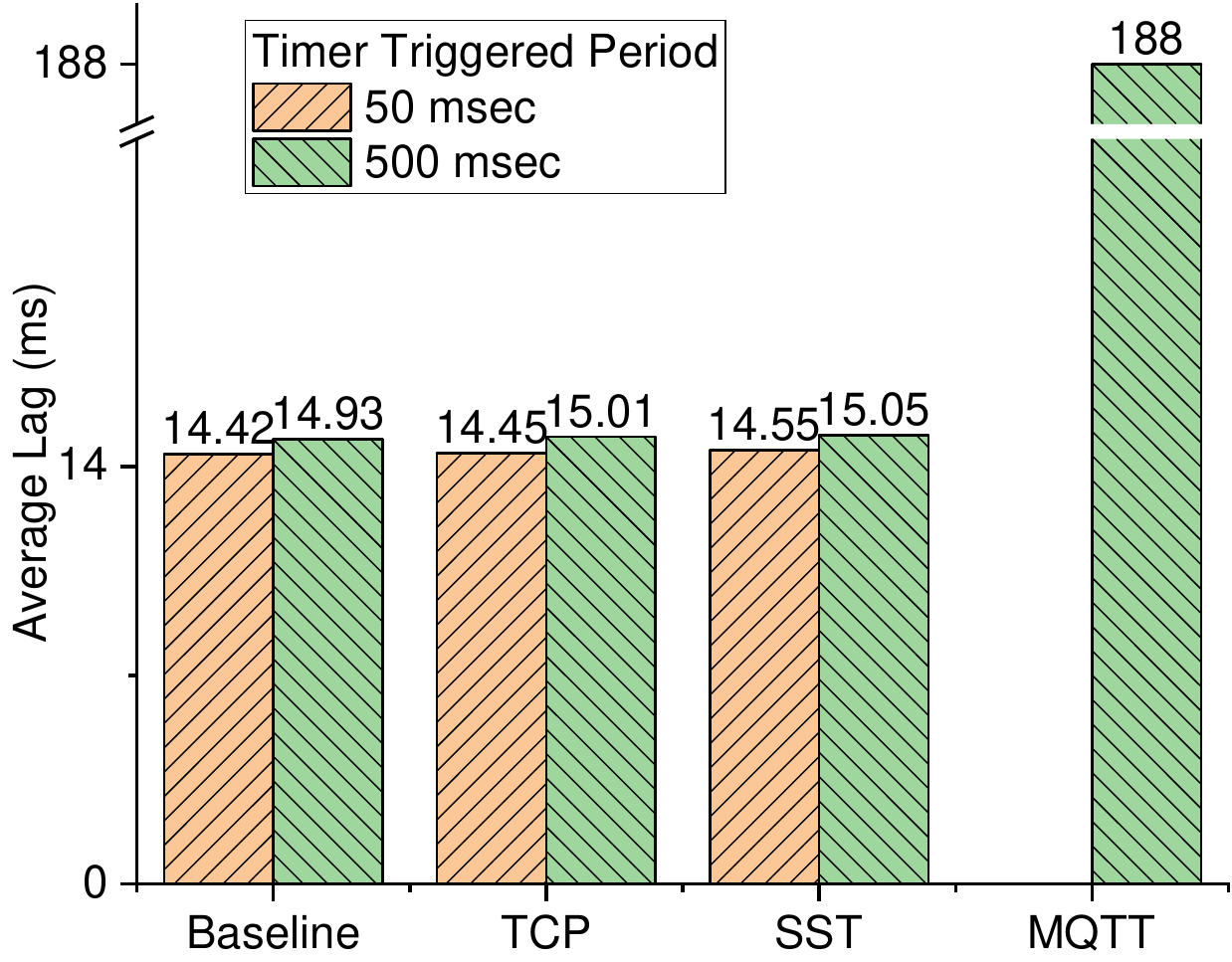}
	\caption{Average lag compared with the baseline LF code. The federate \texttt{Source} sends messages when the timer is triggered as in line~\ref{timer}, with a period of 50 and 500 milliseconds.}
	\label{fig:AverageLag}
\end{figure}

\figurename~\ref{fig:AverageLag} shows the average lag when sending messages in a period of 500 milliseconds and 50 milliseconds.
In the case when the period is 500 milliseconds and the TCP model is compared with the baseline code, there is a 0.53\% increase in lag which is 0.08 milliseconds.
The SST model also has a similar lag of 0.80\% lag increase.
When we compare the SST model to the TCP model, there is only a 0.26\% increase in lag when is security enabled for authentication of federates and confidentiality/integrity of messages.
The 50 millisecond period test also had a similar tendency, showing a very small overhead.

MQTT has a longer lag due to the synchronous behavior.
To ensure deterministic timing, all messages in LF must be sent in order.
MQTT's underlying TCP sends messages in order, but the protocol itself does not guarantee it without QoS level 2.
So, the \textit{netdrv\_write()} must call MQTTClient\_waitForCompletion() after publishing the message to ensure the message is delivered.
For synchronous receive of incoming messages, the \textit{netdrv\_read()} calls MQTTClient\_receive().

As the synchronous parts become the bottleneck, each netdrv\_write() function call takes an average of 90 milliseconds to transfer a message.
Due to the centralized coordination, passing a message via the RTI, the netdrv\_write() is called twice, sending signals from the federate \texttt{Source} to the RTI, and RTI to the federate \texttt{Destination}.
This explains the average latency of 188 milliseconds.
However, when using MQTT, centralized coordination becomes inefficient because the RTI acts as a message broker while MQTT itself has its own broker. 
To remove this inefficiency, LF supports a different mode of runtime execution called \textit{decentralized coordination}, where federates directly communicate with each other without the RTI.
We plan to support this for MQTT in the future.

\subsection{Message Size Overhead}
\begin{figure}
	\centering
	\includegraphics[width=0.7\columnwidth]{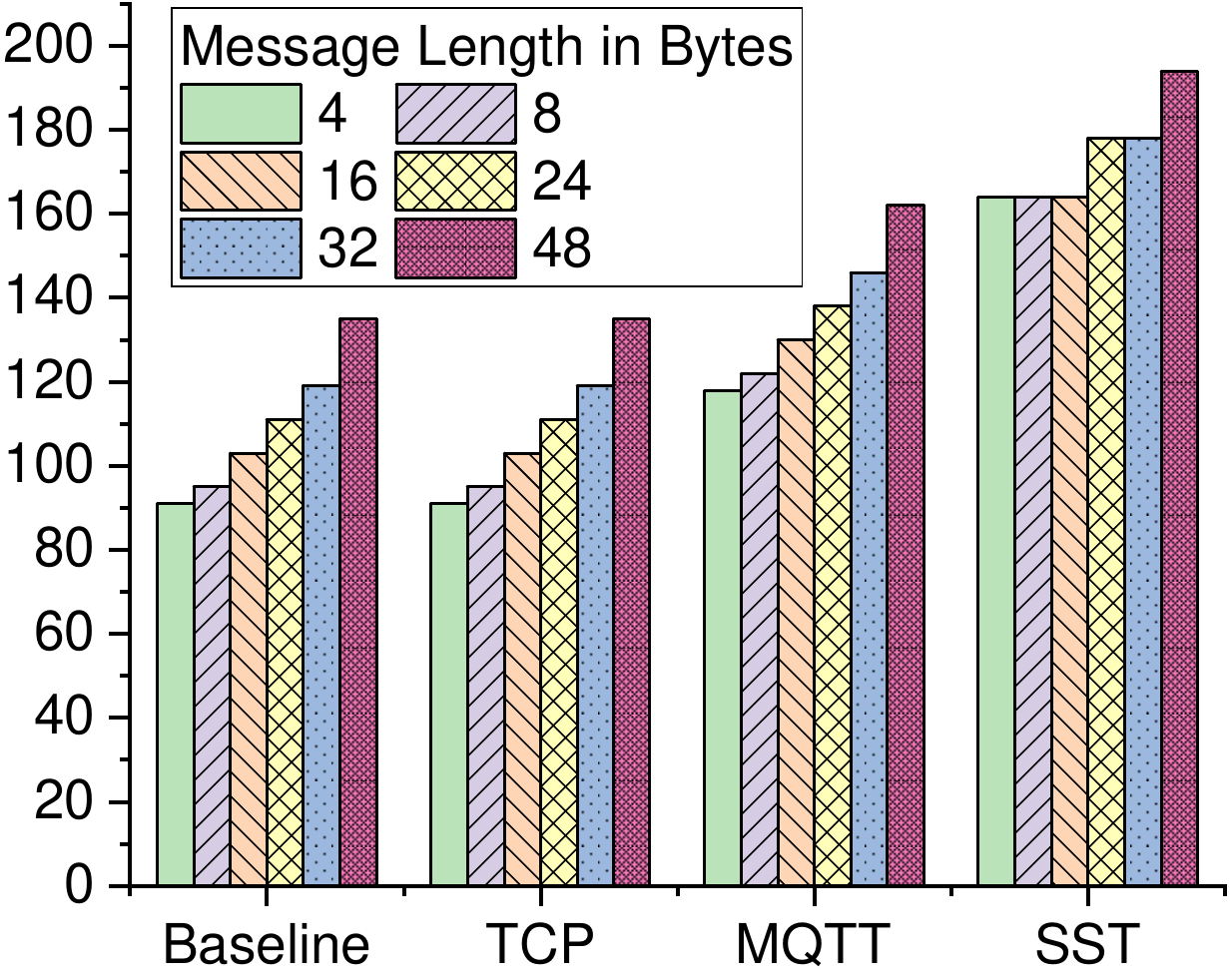}
	\caption{Sizes of messages sent over the network in bytes for different communication types supported by the proposed API compared to the baseline implementation.}
	\label{fig:BytesSent}
\end{figure}
We measure communication overhead in terms of the message size of each communication type using Wireshark~\cite{beale2006wireshark}.
To be specific, we measure the size of MSG\_TYPE\_TAGGED\_MESSAGE, which includes a header with the destination information and a variable payload.
\figurename~\ref{fig:BytesSent} illustrates the total message size when sending 4, 8, 16, 24, 32, and 48-byte long messages as payload.
When this message is sent, a 21-byte header is attached, so when sending a 32-bit integer, a total of 25 bytes are sent.


Details are provided for when 4 bytes are sent.
The baseline code and the client-server model is both based on TCP, so they have the same length of TCP/IP headers of 66 bytes, sending a total of 91 bytes.
The message size of the baseline and client-server models both linearly increase in proportion to the message size.
The MQTT model sent 118 bytes, including 66 bytes of TCP/IP headers and 27 bytes of MQTT headers.
The MQTT header size varies depending on the topic name, which is `MQTTTest\_fed0\_to\_RTI'.

The SST model sent a total of 164 bytes, including the TCP/IP header, SST header, and encrypted message.
SST employs the AES-128-CBC mode, which is a block cipher with a fixed block size of 16 bytes.
As a result, the size of the encrypted output increases in 16-byte increments, demonstrating a stepwise pattern of growth.

Note that the block cipher encryption provided by SST also prevents side-channel attacks.
As described in section~\ref{sec:read/write}, LF has a message type with fixed lengths.
Due to these fixed lengths, eavesdroppers can infer the message type from the message length.
If the eavesdroppers can access to the trace of the messages, the message lengths can be a side channel, allowing them to determine which message types are being transmitted.
By using a block cipher like AES-128-CBC, SST ensures that all encrypted messages are in block sizes, making it difficult to infer the message type.

\subsection{Binary Size Overhead}
\begin{figure}
	\centering
	\includegraphics[width=0.8\columnwidth]{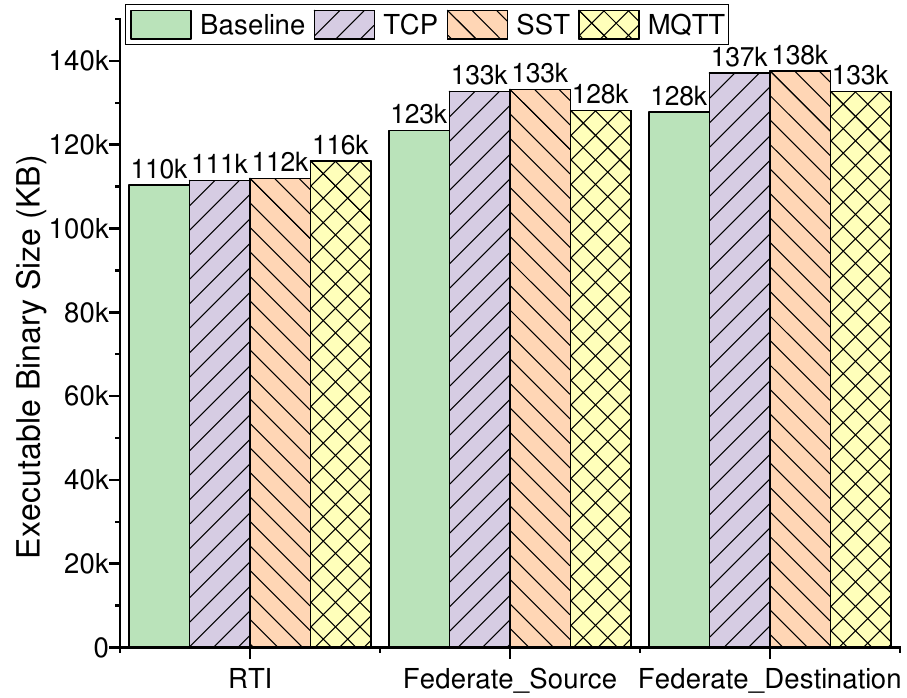}
	\caption{Binary size in kilobytes compared with the baseline LF code.}
	\label{fig:BinarySize}
\end{figure}
To estimate the overhead in terms of binary sizes, we measure the binary size of the RTI, federate \texttt{Source}, and federate \texttt{Destination}.
As illustrated in \figurename~\ref{fig:BinarySize}, the overhead introduced by the abstraction layer is minimal when compared with the baseline code.
The increase in binary size for the RTI is 1.02\%, 5.23\%, and 1.44\% for TCP, MQTT, and SST, respectively.
For the same protocols, the Destination federate's binary size increases by 7.29\%, 3.71\%, and 7.67\%, while the Source federate show a similar increase.

The increased binary size for both RTI and federates mostly comes from compilation of the network abstract layer as a separate library, making it more modular and flexible to extend.
The RTI has a smaller overhead compared to the federates because it functions only as a \textit{Listener}, whereas the federates have dual roles, serving as both \textit{Connectors} and \textit{Listeners} in \textit{decentralized coordination} which has been briefly introduced in Section~\ref{sec:comm_overhead}.
In this case, the RTI does not relay messages to other federates; instead, the federates communicate directly with each other. 
However, the decentralized coordination of currently only supports TCP yet and supporting other protocols will be future work.

Among different modes of communication, MQTT shows the largest binary size for the RTI.
However, for the federates, MQTT has smaller binary sizes than TCP and SST. 
Although we did not perform a detailed analysis of this difference, we speculate compiler optimizations and additional libraries caused this increase.


\subsection{Software Design Analysis}

\begin{figure}
	\centering
	\includegraphics[width=1.0\columnwidth]{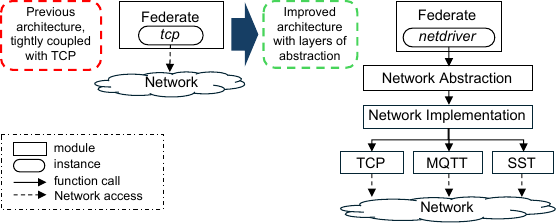}
	\caption{Improved software architecture after applying the proposed API design to the previous LF network node.}
	\label{fig:Decoupling}
\end{figure}

\figurename~\ref{fig:Decoupling} shows the improved software design from the previous existing LF network module.
In the previous software design, as shown in the left side of \figurename~\ref{fig:Decoupling}, each federate owns a TCP socket instance to call network functions to communicate with RTI or other federates.
The direct inclusion of the TCP socket instance requires a large number of code modifications in the federate to add other types of communication protocols (e.g., publish-subscribe or secure communication mode) in addition to the existing TCP protocol.
The right side of \figurename~\ref{fig:Decoupling} denotes the improved software design after applying the proposed API design, which decouples a TCP instance from a federate.

To evaluate the effectiveness of the proposed API in terms of software design, we conduct a qualitative analysis focused on how the application of the proposed approach achieves the software design considerations in ~\cite{hans2008software}.
This paper applies only to qualitative parts of the design considerations, i.e., abstraction, modularity, and information hiding since quantitative evaluation using software engineering metrics is not the primary concern.
Analyzing software design based on quantitative metrics such as cyclomatic complexity~\cite{mccabe1976complexity} is left to future work.

\subsubsection{Abstraction}
The goal of abstraction is to concentrate only on the essential features. To achieve this, two types of abstraction, i.e., \textit{procedural abstraction} and \textit{data abstraction} should be achieved.
\textit{Procedural abstraction} aims to find a hierarchy in the software's control and can be achieved by decomposing the software into sub-modules step-wisely so that the abstraction draws a hierarchical structure.
In this structure, the top node denotes the problem to be solved through the software.
As shown in the right side of \figurename~\ref{fig:Decoupling}, our design shows the APIs to solve the network access from federate as a top node, separates implementation and protocol types as next-level nodes, and forms a hierarchical structure.
Our design also separates protocol types as another layer and hides data types from a federate.

\subsubsection{Modularity}
Coupling between modules should be considered to increase software modularity~\cite{hans2008software}.
Six types of coupling, i.e., content coupling, common coupling, external coupling, control coupling, stamp coupling, and data coupling, need to be analyzed to evaluate the software modularity.
Content coupling occurs when a module changes another module's data or control flows using jump indicators, and common coupling occurs when two modules share data.
Since we separate \textit{Network Abstraction} as the API layer and \textit{Network Implementation} as its implementation, content and common coupling are avoided.
External coupling occurs when modules communicate through an external medium, such as a file, and control coupling occurs when one module sets control flags that are reacted to by the dependent modules.
Neither files nor control flags are used as arguments in our API design.
Either stamp or data coupling occurs by passing complete data structures or simple data between modules.
Our API design also does not pass any data between modules, avoiding all six types of coupling.

\subsubsection{Information Hiding}
The key to implementing information hiding is to hide design decisions and details from other modules.
By designing each selection of a communication protocol as an option \textit{comm-type} in \figurename~\ref{fig:HelloDistributedCode} and implementing each protocol as a separate C source file, we can hide all design decisions and implementation details of each communication protocol into each source file.
This design also enables developers to easily add other communication protocols with a simpler, easier-to-maintain, and more robust interface.

\section{Conclusion and Future Work}

In this paper, we perform a case study of the design of an API and its runtime to achieve network interoperability and security in the IoT and distributed CPS.
The proposed API encapsulates the underlying network implementation layer through seven key API functions. 
We implement our approach using open-source software, Lingua Franca, and SST as a case study.
The evaluation assesses the proposed API design using our case-study implementation by measuring the communication time overhead, message size, and binary size, showing a minimal overhead.
The proposed API and its implementation will be available on GitHub.

As future work, we plan to support more communication modes in our API.
We also plan to allow distributed nodes with different communication modes to join a single federation.
In this case, we envision that a centralized entity such as RTI will be able to handle multiple protocols in one process.
Security options for communication among distributed nodes can be extended to enable fine-grained configurations in our proposed API.

\section*{Acknowledgment}
This work was supported in part by the NSF I/UCRC for Intelligent, Distributed, Embedded
Applications and Systems (IDEAS) and from NSF grant \#2231620.
This work was supported in part by ATTO Research.

%
%
%
 \bibliographystyle{splncs04}
 \bibliography{refs}

%
%
%
%
\end{document}